\documentclass[twocolumn]{aastex62}

\graphicspath{{./}{figures/}}

\shorttitle{DH Tau b rotation rate}
\shortauthors{Xuan}

\begin{document}

\title{A Rotation Rate for the Planetary-Mass Companion DH Tau b}

\correspondingauthor{Jerry W. Xuan}
\email{wxob2015@mymail.pomona.edu}

\author[0000-0002-6618-1137]{Jerry W. Xuan}
\affil{Department of Physics and Astronomy, Pomona College, 333 N. College Way, Claremont, CA 91711}
\affil{Division of Geological and Planetary Sciences, California Institute of Technology, 1200 E. California Blvd., Pasadena, CA 91125, USA}

\author[0000-0002-6076-5967]{Marta L. Bryan}
\affil{Department of Astronomy, University of California, Berkeley, CA 94720, USA}

\author{Heather A. Knutson}
\affil{Division of Geological and Planetary Sciences, California Institute of Technology, 1200 E. California Blvd., Pasadena, CA 91125, USA}

\author[0000-0003-2649-2288]{Brendan P. Bowler}
\affil{Department of Astronomy, The University of Texas at Austin, Austin, TX 78712, USA}

\author[0000-0002-4404-0456]{Caroline V. Morley}
\affil{Department of Astronomy, The University of Texas at Austin, Austin, TX 78712, USA}

\author[0000-0001-5578-1498]{Bj\"{o}rn Benneke}
\affil{D\'{e}partement de Physique, Universit\'{e} de Montr\'{e}al, 2900 Boulevard \'{E}douard-Montpetit, Montreal, Quebec H3T 1J4, Canada}

\begin{abstract}

DH Tau b is a young planetary-mass companion orbiting at a projected separation of 320 AU from its $\sim$2 Myr old host star DH Tau. With an estimated mass of $8-22$ $M_{\rm{Jup}}$ this object straddles the deuterium-burning limit, and might have formed via core or pebble accretion, disk instability, or molecular cloud fragmentation. To shed light on the formation history of DH Tau b, we obtain the first measurement of rotational line broadening for this object using high-resolution (R $\sim$25,000) near-infrared spectroscopy from Keck/NIRSPEC.  We measure a projected rotational velocity ($v$sin$i$) of $9.6\pm0.7$ km/s, corresponding to a rotation rate that is between 9-15\% of DH Tau b's predicted break-up speed. This low rotation rate is in good agreement with scenarios in which magnetic coupling between the companion and its circumplanetary disk during the late stages of accretion reduces angular momentum and regulates spin. We compare the rotation rate of DH Tau b to published values for other planetary-mass objects with masses between $0.3-20$ $M_{\rm{Jup}}$ and find no evidence of a correlation between mass and rotation rate in this mass regime. Finally, we search for evidence of individual molecules in DH Tau b's spectrum and find that it is dominated by CO and H$_2$O, with  no evidence for the presence of CH$_4$. This agrees with expectations given DH Tau b's relatively high effective temperature ($\sim$2300 K).

\end{abstract}

\keywords{exoplanet formation --- exoplanet atmospheres --- high resolution spectroscopy --- brown dwarfs}

\section{Introduction} \label{sec:intro}
Over the past decade, direct imaging searches for self-luminous exoplanets have uncovered a growing number of planetary-mass companions (PMCs) with masses of $5-20$ $M_{\rm{Jup}}$ orbiting at distances of tens to hundreds of AU from their host star \citep[see review by][]{bowler_imaging_2016}. 

There are currently three proposed formation scenarios for PMCs at wide separations: core or pebble accretion, disk instability, and molecular cloud fragmentation. 

The core accretion model \citep{pollack_formation_1996} postulates that giant planets start out by building large solid cores of rocky and icy material which grow large enough to accrete massive gas envelopes. However, low solid densities at the present-day locations of these companions \citep[see review by][]{andrews_mass_2013} mean that the timescale required to grow a core massive enough to undergo runaway gas accretion is expected to be longer than the observed lifetimes of protoplanetary disks. While recent studies have invoked the effects of gas drag on cm-sized solids (pebble accretion) in order to grow solid cores faster, these pebbles also undergo relatively rapid radial migration, reducing the pebble surface density in outer regions of the disk \citep[e.g.,][]{rosenthal_restrictions_2018, lin_balanced_2018}. 

In models of disk instability, companions form rapidly through local gravitational collapse in a protoplanetary disk \citep{boss_giant_1997,boss_formation_2006,Dodson-Robinson2009,vorobyov_formation_2013}. However, disk surface densities tend to be too low for gravitational instability to operate at separations beyond $100$ AU \citep{Dodson-Robinson2009}. While it has been proposed that these companions could have formed closer to their host stars and been subsequently scattered out to their present day locations by a more massive body in the system, scattering is unlikely to be a dominant formation pathway for this population of companions \citep{bryan_searching_2016}. Alternatively, these PMCs could have formed through the fragmentation of a molecular cloud in a process akin to stellar binary formation \citep{bate_formation_2002}. However, hydrodynamical simulations have trouble explaining the extreme mass ratios (a few percent) of the observed population of PMCs \citep{bate_stellar_2012}.

Previous studies have investigated the origins of these wide-separation planetary-mass companions by examining their mass and semi-major axis distributions \citep{brandt_statistical_2014, nielsen_gemini_2019, wagner_mass_2019}. Most recently, \citet{nielsen_gemini_2019} reported results from a 300-star survey using the Gemini Planet Imager \citep{Macintosh2014}. In this study, they found tentative evidence that planetary-mass companions ($2-13$ $M_{\rm{Jup}}$) have power law distributions in mass and semi-major axis that are distinct from those of brown dwarf companions. However, this study was based on a total of nine companions (six planets and three brown dwarfs) with projected separations less than 60 AU, and was therefore limited in its statistical leverage. 

Previous studies have also searched for evidence of a correlation between planet occurrence rate and stellar metallicity in order to distinguish between core accretion and other formation mechanisms. There is compelling evidence to suggest that relatively close-in ($<$10 AU) and low-mass ($<$10 $M_{\rm{Jup}}$) gas giant planets likely form via core accretion, as they are preferentially found around more metal-rich stars \citep{fischer_planet-metallicity_2005, schlaufman_evidence_2018}. This metallicity correlation disappears for transiting planets larger than $\sim$8 $M_{\rm{Jup}}$, indicating that more massive companions may form via an alternative mechanism, most likely gravitational instability \citep{schlaufman_evidence_2018}. Indeed, close equal-mass stellar binaries (semi-major axes less than 10 AU) preferentially occur in low metallicity environments \citep{moe_close_2019, el-badry_discovery_2019}, suggesting that low disk metallicities do in fact favor gravitational instability mechanisms. 

The atmospheric compositions of individual directly imaged planets can also be used to place constraints on their formation and migration histories  \citep{konopacky_detection_2013, barman_simultaneous_2015}. While we expect disk instability and molecular cloud fragmentation to produce companions with stellar atmospheric composition, core accretion is expected to produce companions with non-stellar atmospheric compositions \citep{oberg_effects_2011, espinoza_metal_2017}.  However, these objects have complex atmospheric chemistries that are additionally altered by the formation of condensate cloud layers \citep[e.g.,][]{line_uniform_2015, burningham_retrieval_2017}, and it is therefore difficult to obtain reliable atmospheric abundances from the current body of low- and medium-resolution spectra available for these objects.

In this study we focus instead on rotation rates as probes of the formation and accretion histories of these objects. Independent of formation mechanism, accreting protoplanets are expected to form circumplanetary gas disks that transfer angular momentum to the planet \citep{ward_circumplanetary_2010}, causing the planet to spin up. After the circumplanetary disk is dispersed, planets cool down and contract in size, thereby spinning up further. Without any braking mechanism, young accreting planets should spin up to speeds approaching the break-up velocity. However, Jupiter and Saturn both rotate $3-4$ times slower than their breakup velocities, suggesting that some mechanism(s) helped regulate their spins. 

\citet{takata_despin_1996} first suggested that the hydromagnetic torque arising from the interaction between the planet's magnetic field and the partially ionized circumplanetary disk could dissipate enough angular momentum to account for the present day spins of the solar system gas giants. More recently, \citet{batygin_terminal_2018} developed a new model which demonstrated that effective magnetic coupling between a slower rotating disk and a faster rotating planet could dissipate enough angular momentum to be consistent with the spin measurements made in \citet{bryan_constraints_2018}.

Given this general picture of spin regulation, it is reasonable to expect that variations in circumplanetary disk properties resulting from these three formation models \citep{szulagyi_circumplanetary_2017} could lead to variations in an object's final rotation rate. With this idea in mind, \citet{bryan_constraints_2018} compared the rotation rates of five planetary-mass companions with those of six free-floating, planetary-mass brown dwarfs, which formed via molecular cloud fragmentation. They found that the spin distributions between these two populations were consistent, suggesting either that the formation mechanism does not alter circumplanetary disk properties in a way that is relevant for spin-down, or that both PMCs and isolated planetary-mass brown dwarfs formed via the same mechanism. All of the objects rotated at just 10-20\% of their break-up velocities regardless of age, consistent with a picture in which planetary spin is set by the time the gas disk dissipates, likely through interactions between the planet and its circumplanetary disk. 

In this study, we measure a rotation rate for the young, wide separation planetary-mass companion DH Tau b. This object was discovered by \citet{itoh_young_2005} with CIAO/Subaru and straddles the deuterium burning limit with an estimated mass of $8-22$ $M_{\rm{Jup}}$ \citep{luhman_discovery_2006}. It orbits at a projected separation of 320 AU\footnote{Based on parallax measurements from Gaia DR2, and astrometry measurements from \citet{bryan_searching_2016}.} from the 2 Myr old T Tauri star DH Tau (0.64$\pm$0.04 $M$\textsubscript{\(\odot\)}), which is itself part of an ultra-wide binary (2210 AU) with DI Tau \citep{Kraus_Hillenbrand_2009}. Assuming literature mass estimates, the mass ratio for the DH Tau system is between 1.1-3.5\%.

In \S\ref{sec:data}, we describe our observations of the DH Tau system with Keck/NIRSPEC and our spectral extraction pipeline. In \S\ref{sec:methods} we detail our measurement of the projected rotation rate ($v$sin$i$) of the companion. In \S\ref{sec:discuss}, we discuss how this rotation rate measurement fits into the context of previously established correlations between rotation rate, mass, and age. We also search for evidence of individual molecules in the atmosphere of DH Tau b. Finally, we summarize our conclusions in \S\ref{sec:conclude}.

\section{Observations and Spectral Extraction} \label{sec:data}
We obtained $K$ band (2.03-2.38~$\mu$m) spectra of both DH Tau b and DH Tau using the near-infrared high-resolution (R$\sim$25,000) spectrograph NIRSPEC at the Keck II telescope on UT November 3 2017. We used NIRSPEC in AO mode with a $0.041\times2.26^{\prime\prime}$ slit and targeted the companion and its host star separately since their angular separation (2.3\arcsec) was larger than the length of the slit. Since the predicted stellar contrast at the location of the companion (40 resolution elements in $K$ band, corresponding to $\Delta$mag $\approx$ 15.5) is much larger than the intrinsic companion-to-star contrast ($\Delta$mag $\approx$ 5.9 as measured by \citealt{bryan_searching_2016}) in $K$ band, we estimate that star should contribute less than 0.02\% of the flux in the companion aperture. For DH Tau b we adopted an AB nod pattern with eight nods and used an integration time of 900 seconds for each image, which amounts to a total integration time of two hours. For DH Tau we performed a single ABBA nod sequence, with a total integration time of sixty seconds.

\begin{figure}[t]
    \centering
    \includegraphics[width=0.9\linewidth]{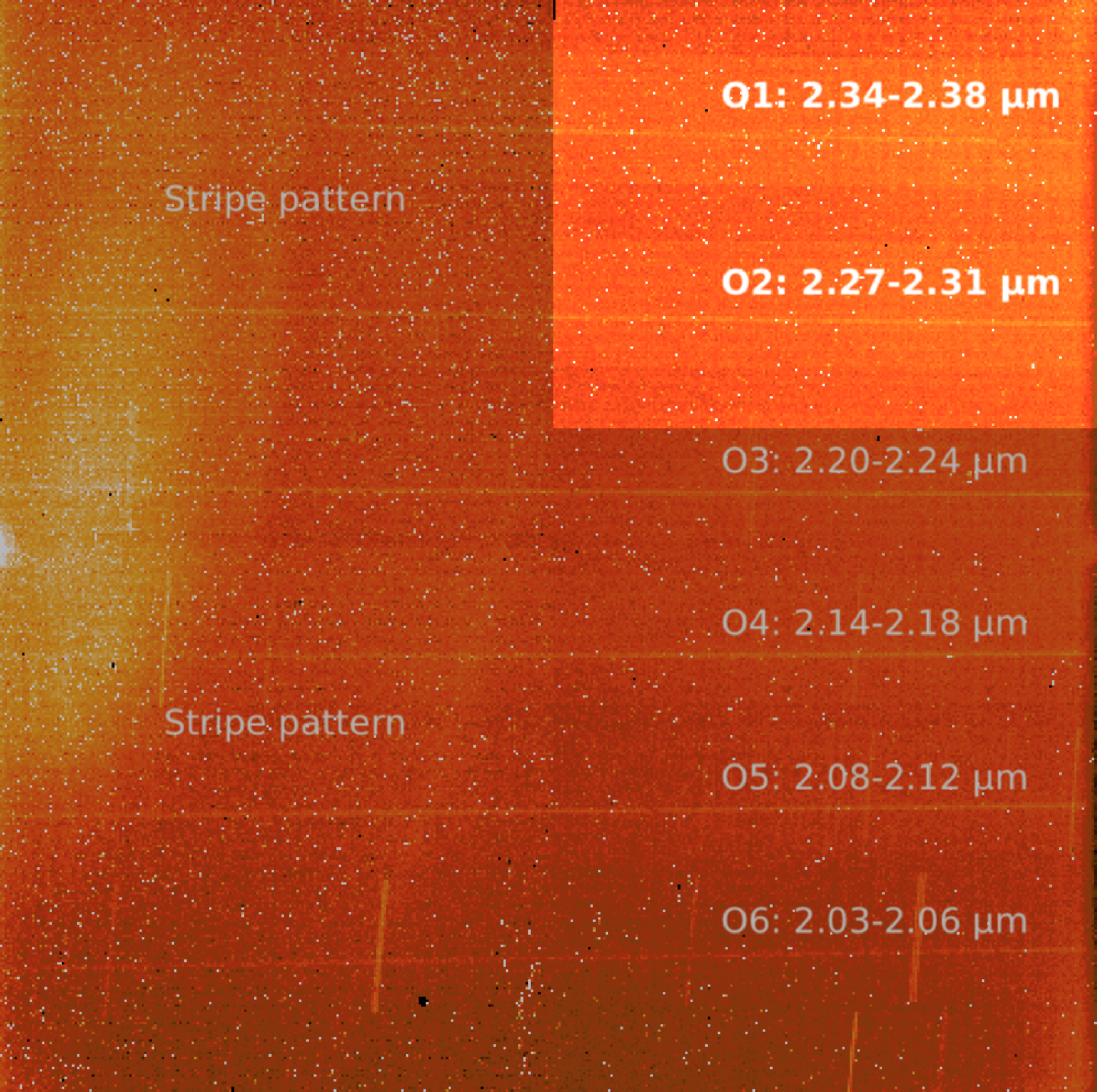}
    \caption{A raw image for the spectrum of DH Tau b. The six spectral orders are distributed vertically, with the portions of the spectrum used in our final rotation rate analysis highlighted.
    \label{fig:raw_data}}
\end{figure}

The resulting raw data consists of a series of $1024\times1024$ pixel images (see Fig.~\ref{fig:raw_data}). Each image contains six spectral orders that cover the $K$ band wavelength range ($2.03-2.38$ $\mu$m). Within each order, wavelength varies along the x axis. As part of pre-processing, we flat-field and remove bad pixels from our raw AB images. We then difference the AB nod pairs in order to subtract out the sky background and dark current, yielding one positive and one negative trace for each spectrum. Finally, we median-combine the different sets of A$-$B images order by order, and work with these median-combined orders in the subsequent steps.

We find that our raw data exhibit a stripe pattern in the two left quadrants (visible in Fig.~\ref{fig:raw_data}). This effect was noted by \citet{bryan_constraints_2018}, who attributed the stripes to bias voltage variations in the NIRSPEC detector. We correct for this effect by calculating the median value of the unaffected rows and adding or subtracting a constant value from the striped rows to match this value. This correction reduced the amplitude of the stripe pattern in the raw images, but we found that the resultant 1D spectrum was ultimately too noisy to include in the final measurement. We therefore opt to discard the left half (short-wavelength half) of each spectral trace.

\begin{figure*}
    \centering
    \epsscale{1.1}
    \plottwo{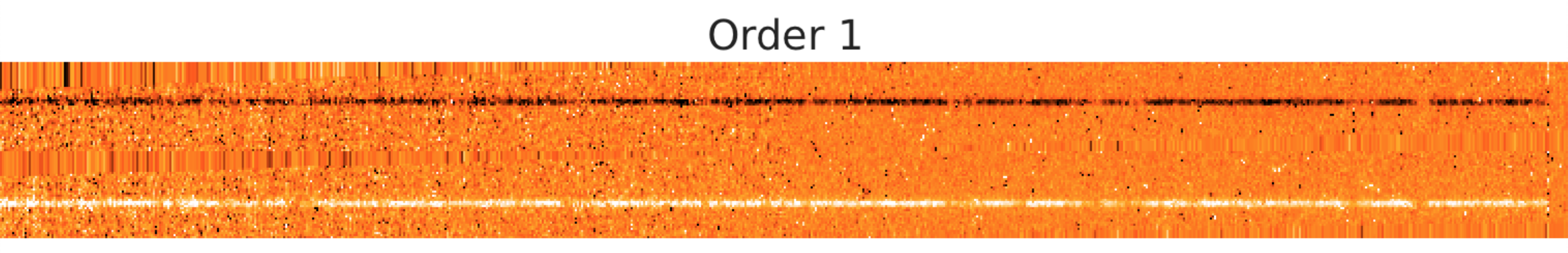}{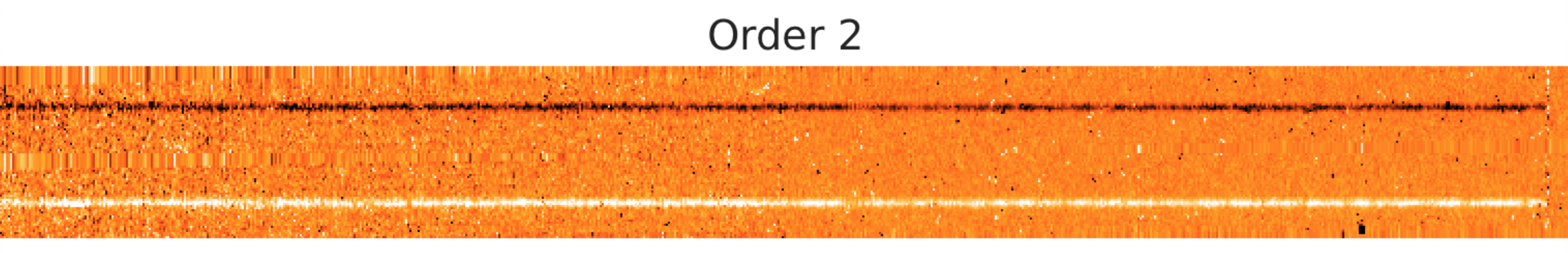}
    \caption{The combined and rectified A$-$B image for DH Tau b order 1 and 2 observations. After rectifying, the traces lie along the $x$ direction, allowing us to convert them into 1D spectra by integrating along the $y$ (cross-dispersion) direction.
    \label{fig:A_B_image}}
\end{figure*}

\subsection{1D Extraction and Wavelength Calibration} \label{sec:rectify}
Given the modest curvature of the NIRSPEC traces, we rectify each median-combined order by determining the vertical ($y$) position of the trace within each column and then fitting a third order polynomial to these positions. To estimate the vertical position of the trace in each column, we fit Gaussian functions along the trace and perform checks on the width, height, and location of the fitted Gaussian functions. We then resample each trace onto a rectified grid using linear interpolation (Fig.~\ref{fig:A_B_image}).

After generating a 2D rectified spectrum for each median-combined order, we extract the 1D spectrum by summing in the $y$ (cross-dispersion) direction using optimal estimation \citep{horne_optimal_1986}. In summary, for each median-combined order we calculate an empirical point spread function (PSF) profile at each $x$ position along the cross-dispersion ($y$) axis using the median of flux values. We then use the PSF profile at each $x$ (wavelength) position to take the weighted sum of the flux in the $y$ direction, where the optimal weights are given by the square of the PSF profile divided by the variance of the flux, as derived by \citet{horne_optimal_1986}. This procedure collapses the 2D spectral trace into a 1D spectrum.

We next convert our 1D spectrum in pixel space to wavelength space. We determine the wavelength solution using the telluric lines imprinted on the stellar spectrum, since the stellar spectrum is much brighter than that of the companion. Specifically, we create telluric models with the radiative transfer code RFM \citep{dudhia_reference_2017} and fit them to the stellar spectrum assuming the wavelength solution is a fourth-order polynomial function of the pixel position. As we expect the wavelength solution for both objects to be the same except for a linear offset, we apply the wavelength solution from the star to the companion, and fit an additional linear offset term to account for the different placements of the two targets within the slit. We calculate this linear offset with a cross-correlation method, sliding the companion spectrum in wavelength space and searching for the offset position where the telluric lines in the companion spectrum best match the telluric model in the corresponding wavelength range.

\subsection{Telluric Removal}
\begin{figure*}[t]
    \centering
    \includegraphics[width=0.45\linewidth]{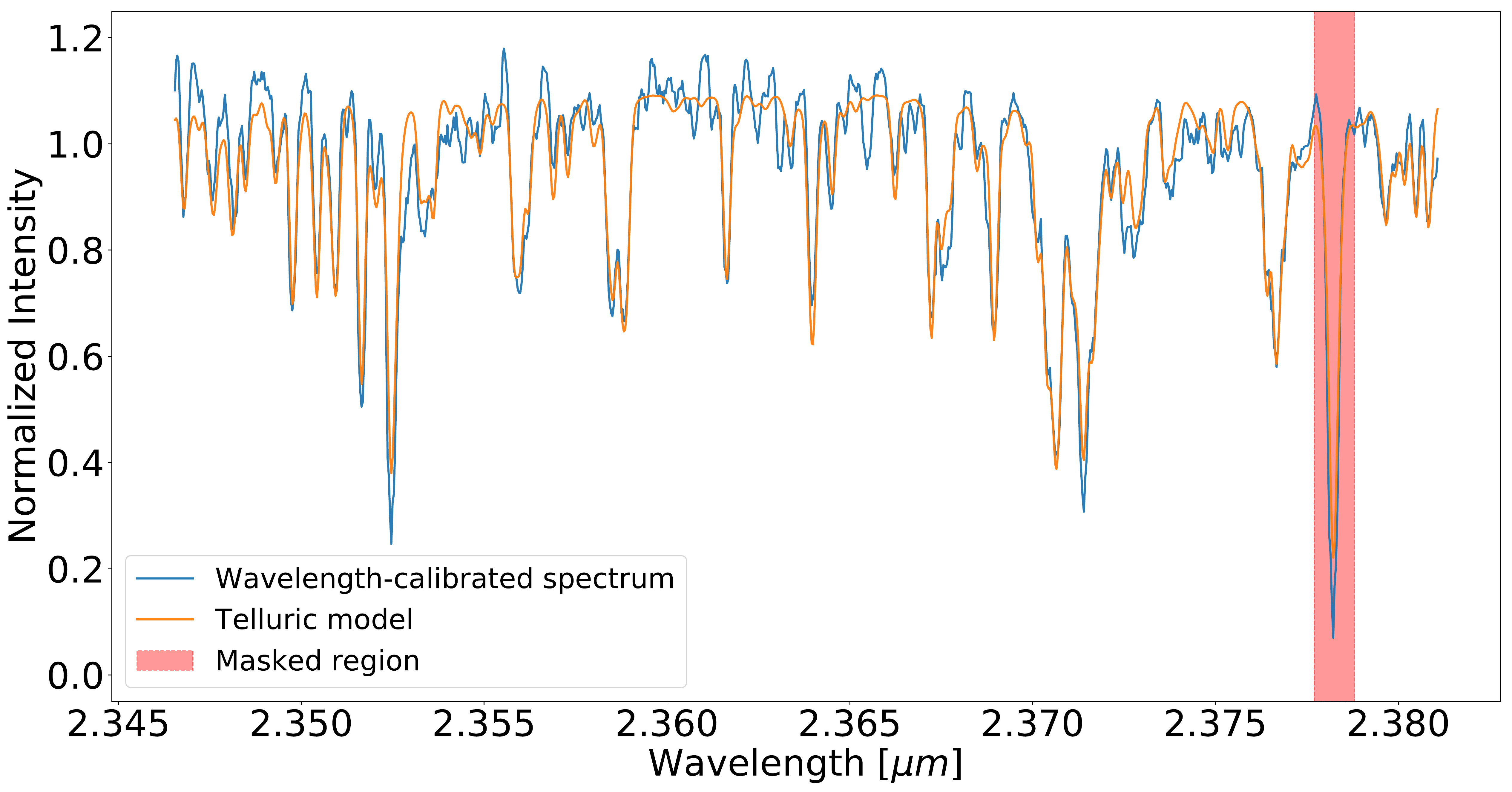}
    \includegraphics[width=0.45\linewidth]{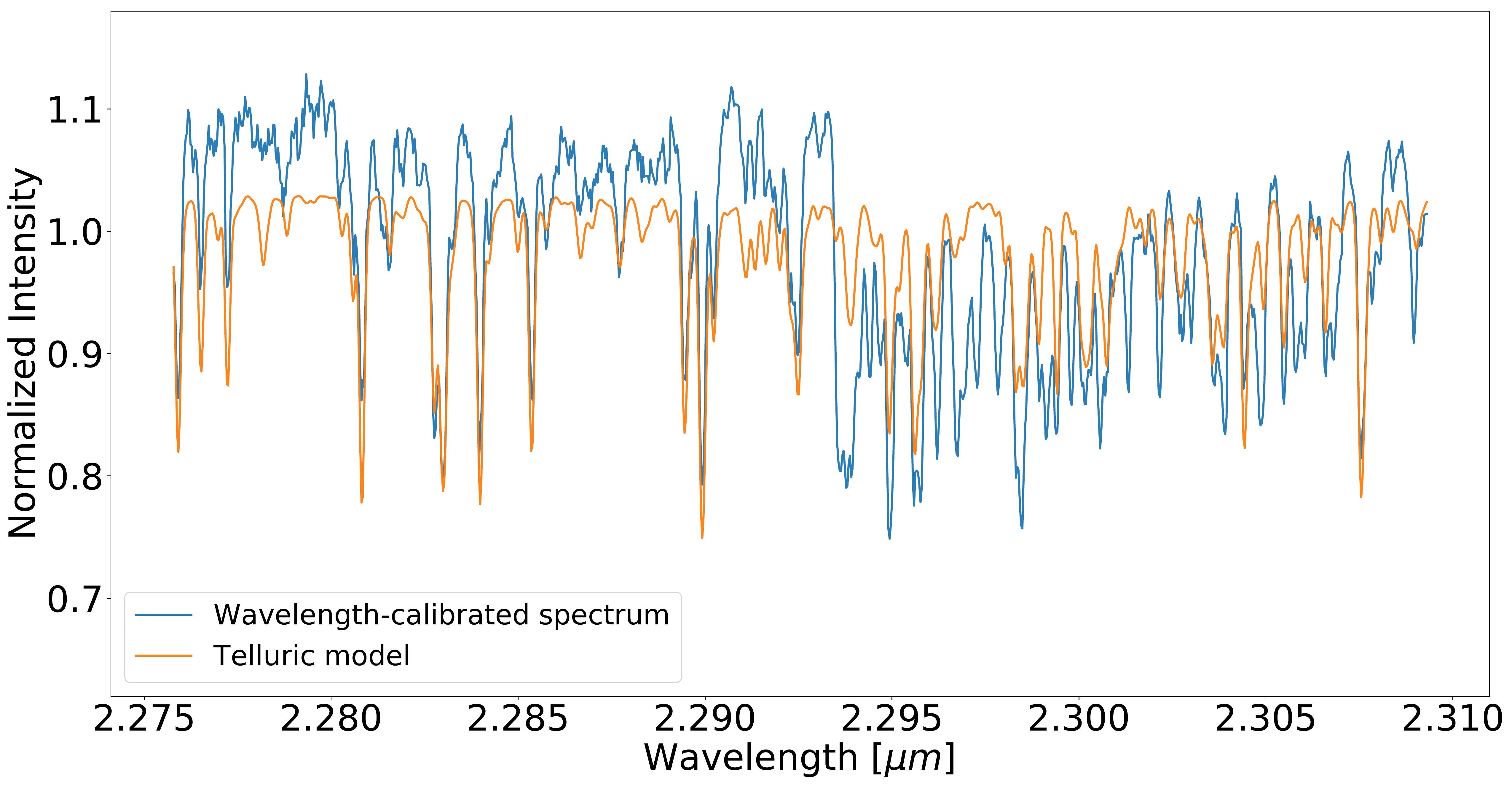}
    \caption{Best fit telluric models in orange and wavelength-calibrated spectra in blue for order 1 (left) and order 2 (right) of DH Tau. We mask out the deepest telluric line in order 1 as it creates a substantial artifact in the telluric-corrected spectrum. \label{fig:molecfit}}
\end{figure*}

After wavelength-calibrating both the stellar and companion spectra, we fit a new set of telluric models to the spectra order-by-order to remove the telluric signal from the data. For this step, we use the software \texttt{molecfit} \citep{smette_molecfit:_2015, kausch_molecfit:_2015}, which uses the radiative transfer code Line-by-line Radiative Transfer Model (LBLRTM). We use \texttt{molecfit} to empirically fit telluric models from our spectra, varying the molecular abundances and instrumental resolution (modeled by a single Gaussian function) to find the best fit (see Fig.~\ref{fig:molecfit} for an example). In addition, we use \texttt{molecfit} to perform an iterative continuum fit (with a third order polynomial) to flatten out the black-body continuum in the spectrum, and fine-tune our wavelength solution using another fifth order polynomial.

We divide out the best-fit telluric model from each order to produce a telluric-corrected spectrum. However, the telluric models are an imperfect match, leaving artifacts in the corrected spectrum. These artifacts are most pronounced around deep lines where there is a mismatch in line shape between model and data. Order 1 has several deep telluric absorption lines. We mask the deepest of these lines, centered at $\sim$2.3782 $\mu$m. The total flux in this line is less than 23\% of the continuum value and it therefore contains little useful information.

\subsection{Selection of Spectral Orders} \label{sec:selection}
\begin{figure*}[t]
    \centering
    \plottwo{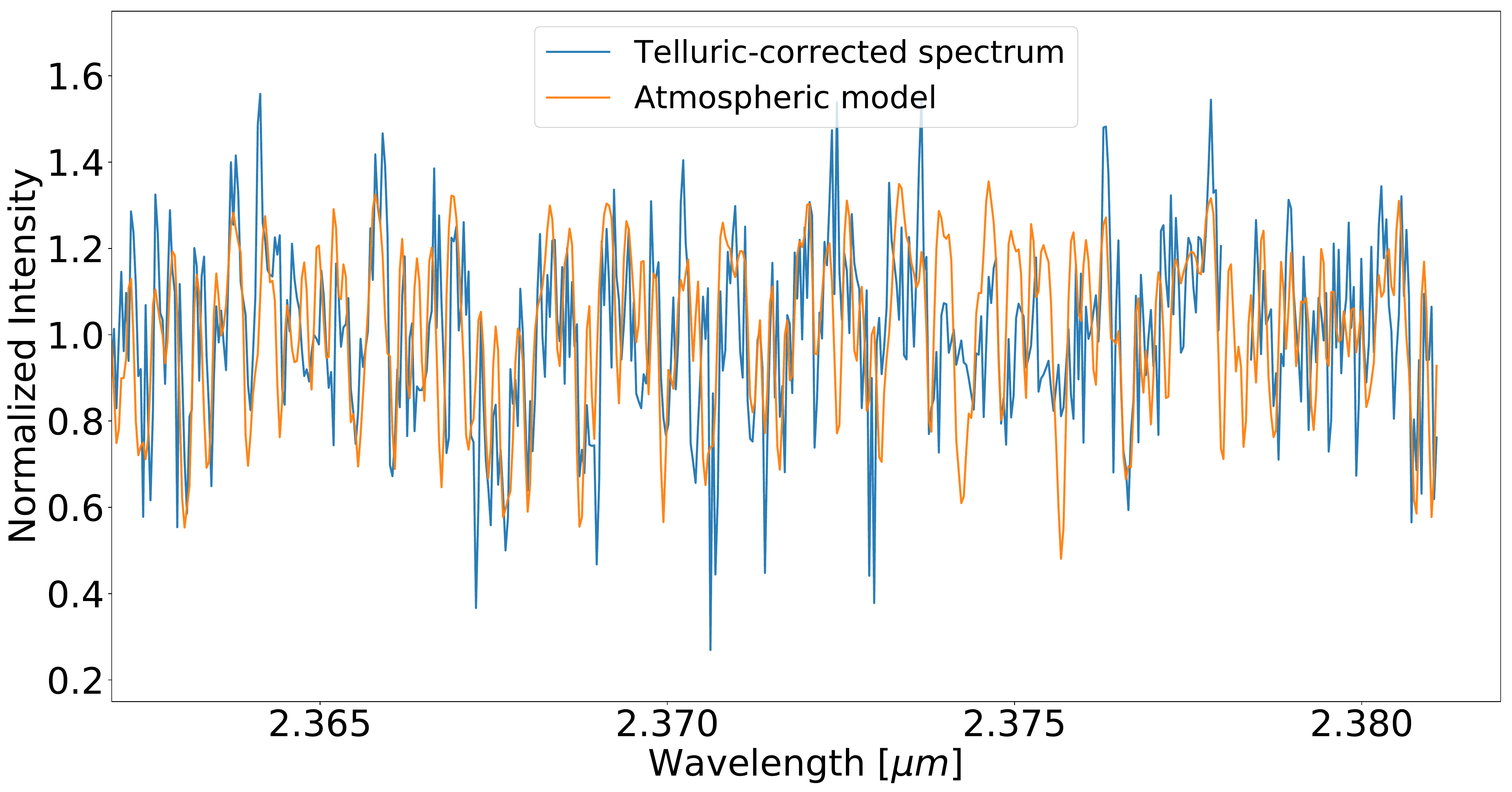}{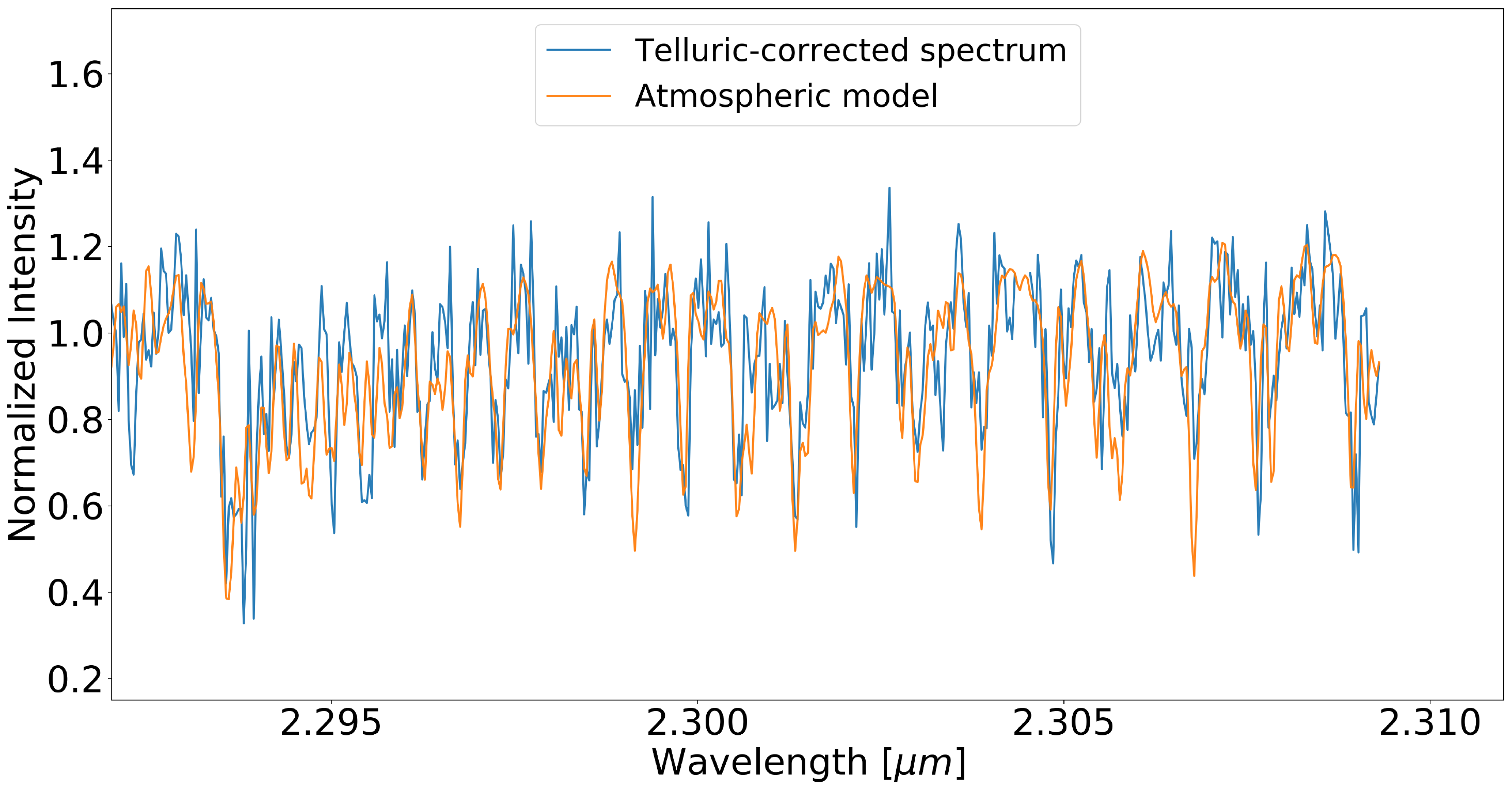}
    \caption{Telluric-corrected spectrum (blue) for DH Tau b order 1 (left panel) and order 2 (right panel). Overplotted in orange are the corresponding model Sonora spectra with parameters that match those of the companion. We use these data and model pairs to calculate the cross-correlation function and measure rotational line broadening.
    \label{fig:tel_cor_spec}}
\end{figure*}

We perform the reduction process for both the host star and the companion, and obtain reduced spectra (wavelength-calibrated and telluric-removed) for both objects (see Fig. \ref{fig:tel_cor_spec} for an example). For our subsequent analysis, we utilize two out of the six spectral orders: order 1 ($2.34-2.38$ $\mu$m) and order 2 ($2.27-2.31$ $\mu$m). We find that these orders have the most accurate wavelength solutions and therefore typically have cleaner telluric corrections than the other four orders. The wavelength solutions for the discarded orders tend to be inaccurate because they contain relatively few telluric lines. Orders 1 and 2 also span prominent absorption lines from carbon monoxide and water in the planet's spectrum, making them some of the most information-rich orders for measuring rotational line broadening. For consistency, we also limit our fits to the stellar spectra to these same two orders.

\section{Spectral Analysis}\label{sec:methods}
\subsection{Instrumental Resolution} \label{sec:res}
The instrumental broadening ($R$ = $\lambda$/$\Delta\lambda$) is degenerate with rotational line broadening, so our ability to measure $v$sin$i$ depends on the accuracy of our value for $R$.  The instrumental resolution is also used as an input for broadening the telluric models. We estimate $R$ using two independent methods and confirm that they agree with each other. Our primary measurement comes from \texttt{molecfit}, which fits single Gaussian functions to the telluric lines in a spectrum in order to measure their width. We use \texttt{molecfit} to fit the telluric lines in the wavelength-calibrated (prior to telluric removal) companion spectrum, where we allow the instrumental resolution $R$ to vary separately for each order and each trace.\footnote{We do not use arc lamp images to measure $R$ since it is possible that the object size is smaller than the slit width during our observation. The arc lamp would merely measure the resolution corresponding to the slit width, which could be lower than the actual resolution.} \texttt{Molecfit} returns the Gaussian FWHM in pixels, which we convert to resolution $R$ using  $\lambda$/$\Delta\lambda$ where $\lambda$ is the mid-wavelength of the given spectral order and $\Delta\lambda$ is the minimum difference in wavelength that is resolvable by the instrument.

We obtain four independent estimates for $R$ from orders 1 and 2 of DH Tau b, where each order contains two traces (positive and negative). All measurements except the negative trace of order 2 agree with each other within $3\sigma$ (see Table~\ref{tab:R_estimates} for a summary). {We found that the spectrum from this outlier trace contains many narrow noise features that skew the resolution estimate by forcing the fit to use broad lines (i.e., lower resolution) to encompass a series of narrow spikes. Indeed, this trace prefers a much smaller value for $R$ that is inconsistent with the width of the NIRSPEC slit we used ($0.041^{\prime\prime}$). This slit width is sampled by 3 pixels on the detector, setting a maximum resolution element of 3 pixels (when the PSF of the target fills the entire slit), whereas the measurement of $R$ from this trace corresponds to a 5-pixel wide slit (which is non-physical). We thus discard the estimate from the negative trace of order 2. 

While resolution varies as a function of wavelength, we estimate that the average resolution varies by only $\sim$2\% between orders 1 and 2 due to the difference in central wavelength and dispersion. Because the size of the wavelength-dependent effect on $R$ is less than $\sim$0.3$\sigma$ (see Table~\ref{tab:R_estimates}), we can ignore this effect. Taking the weighted average of the three remaining measurements, we estimate that $R$ = 24,800$\pm$1000.

\begin{deluxetable}{ccccc}
\tablecolumns{5}
\tabletypesize{\small}
\tablewidth{\textwidth}
\tablecaption{\normalsize{\texttt{Molecfit} resolution estimates for DH Tau b} \label{tab:R_estimates}}
\tablehead{\colhead{Order} & \colhead{FWHM} & \colhead{FWHM} & \colhead{$R$} & \colhead{$R$}\vspace{-8pt} \\
\colhead{number} & \colhead{(Pos)} & \colhead{(Neg)} & \colhead{(Pos)} & \colhead{(Neg)}}
\startdata
1 & 2.82$\pm$0.17 & 2.68$\pm$0.18 & 23,500$\pm$1,400 & 24,800$\pm$1,700\vspace{+2pt} \\
2 & 2.34$\pm$0.19 & 5.01$\pm$0.44 & 28,800$\pm$2,500 & 13,500$\pm$1,200\vspace{+2pt} \\
\enddata
\tablecomments{Estimates of instrumental resolution from telluric fits using \texttt{molecfit} for the positive and negative traces of the two orders used in this study. We report 1$\sigma$ errors computed from the covariance matrix. Resolution is expressed in terms of the pixel size of the Gaussian FWHM (left two columns), and resolving power $R$ (right two columns). We exclude the estimate from the negative trace of order 2 in our weighted average, as the derived $R$ is unphysically low given the width of the NIRSPEC slit used for these observations.}
\end{deluxetable}

Because we know the rotation rate of DH Tau from a previous study \citep{nguyen_close_2012}, we can obtain an independent estimate of the instrumental broadening by fitting the total amount of line broadening in the stellar spectrum. For this fit we use a model stellar spectrum from the PHOENIX spectral library \citep{husser_new_2013}, assuming $T_{\rm{eff}}$=3700 K \citep{andrews_mass_2013}, log($g$)=3.50 m$\rm{s}^{-2}$, and solar metallicity. We find that this second approach yields $R$ = 25,200$\pm$2,700, which is consistent with our first measurement albeit with significantly larger uncertainties. We therefore adopt a value of $R$ = 24,800$\pm$1000 based on the telluric fits in the subsequent analyses.

\subsection{Rotational Line Broadening} \label{sec:mcmc}
We measure rotational line broadening for DH Tau b using a cloud-free Sonora model with effective temperature, surface gravity, and metallicity set to the estimated values for DH Tau b. The Sonora model atmosphere grid is appropriate for the atmospheres of brown dwarfs and young giant planets \citep{marley_sonora_2018}; the models are available online\footnote{https://doi.org/10.5281/zenodo.2628068
}. The Sonora models are calculated using methods that are extensively described in \citet{mckay_thermal_1989}, \citet{marley_atmospheric_1996}, \citet{Marley_2002}, \citet{Saumon_2008}, \citet{Morley_2012}, and \citet{Morley_2014}. The opacity database for gases is described in \citet{Freedman_2008} and \citet{Freedman_2014}. Updates were made to the opacities of a number of species, including the alkali metals, water, and methane. The abundances of molecular, atomic, and ionic species are calculated using a modified version of the NASA CEA Gibbs minimization code \citep{McBride_92}. Further details on the opacities and chemical equilibrium are described in Marley et al. (in prep.).

We use COND models from \citet{baraffe_evolutionary_2003} to estimate $T_{\rm{eff}}$ and log($g$) for DH Tau b by inputting measured values of log($L_{\rm{Bol}}$/$L$\textsubscript{\(\odot\)}) and age. We adopt an age of 2$\pm$1 Myr, corresponding to the median age of Taurus \citep{bertout_evolution_2007}, a log($L_{\rm{Bol}}$/$L$\textsubscript{\(\odot\)}) of -2.71$\pm$0.12 as measured by \citet{luhman_discovery_2006}, and assume a solar metallicity. This gives us an estimated $T_{\rm{eff}}$ of 2300$\pm$100 K and log($g$) of 3.7$\pm$0.1 dex, in good agreement with the values reported in \citet{bonnefoy_library_2014}. We utilize our independently derived values for $T_{\rm{eff}}$ and log($g$) in this study in order to facilitate comparisons with the objects described in \cite{bryan_constraints_2018}. This is particularly important when calculating the predicted break-up velocity for each object, which we use as a normalization factor on the projected rotation rate. 

\begin{figure}[t]
    \centering
    \includegraphics[width=\linewidth]{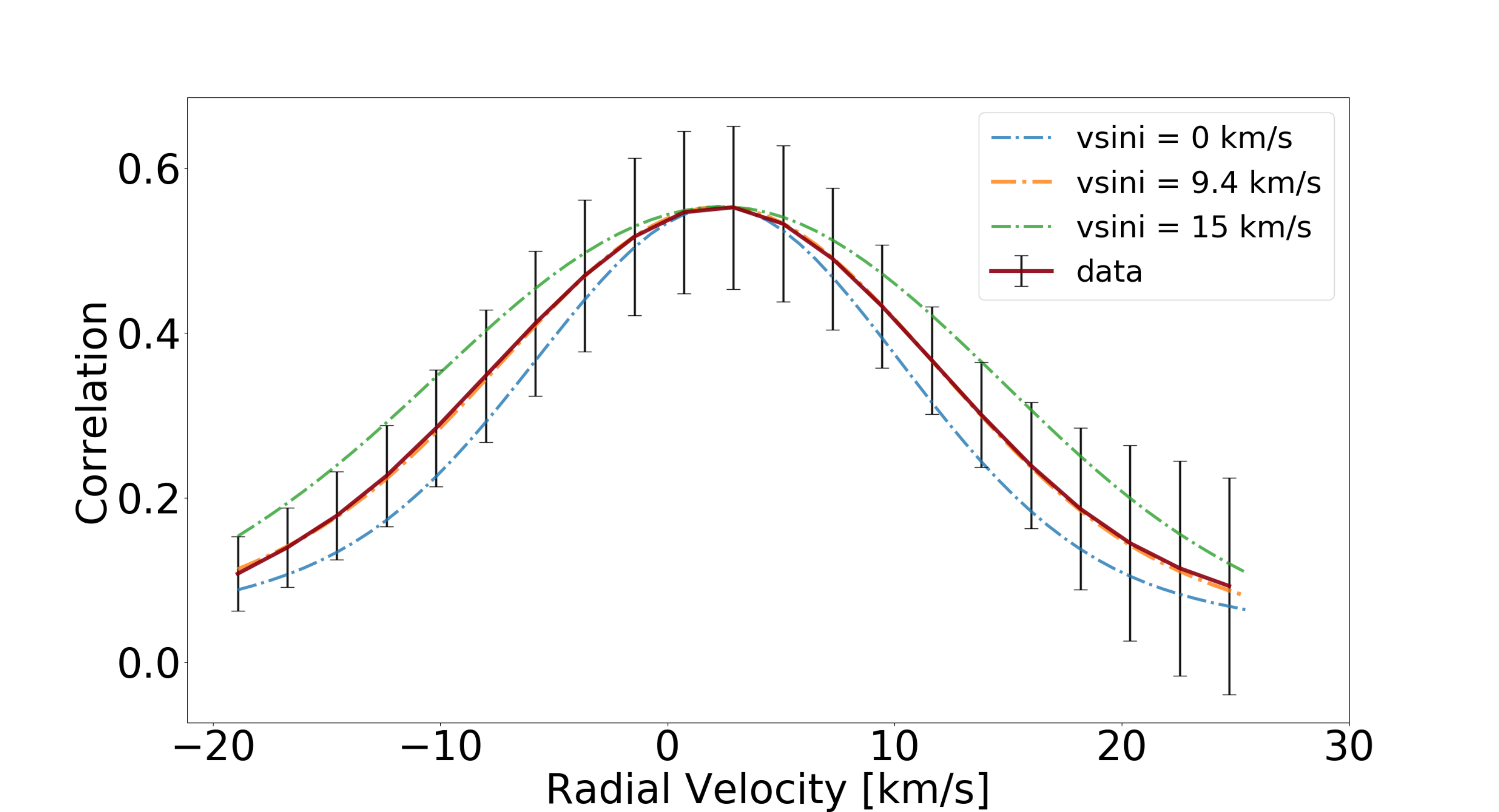}
    \caption{Cross-correlation between the observed spectrum of DH Tau b (order 2 negative trace) and a model Sonora spectrum broadened by the instrumental resolution $R$ (red line). Error bars are generated using a jackknife resampling technique (see Eq.\ref{eq:jackknife}). We also show representative model CCFs between a model spectrum broadened to $R$ and the same model additionally broadened by a range of $v$sin$i$ values (colored lines). The best-fit CCF has a $v$sin$i$ of 9.4 km/s, and a RV value corresponding to the x position of the CCF peak.
    \label{fig:ccfs}}
\end{figure}

\begin{figure}[t]
    \centering
    \includegraphics[width=\linewidth]{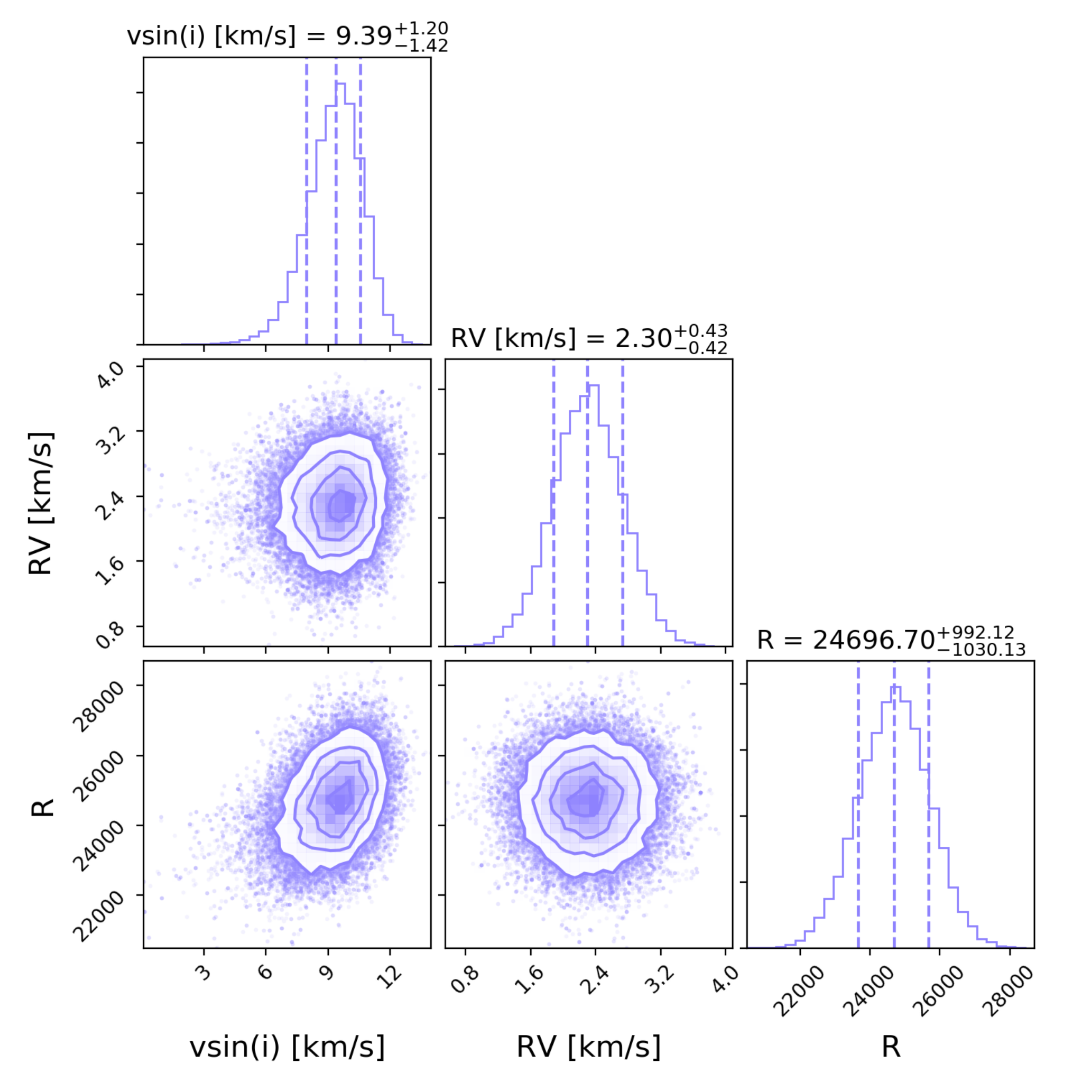}
    \caption{Posterior probability distributions for $v$sin$i$, RV, and log($R$) from our fit to the negative order 2 trace for DH Tau b. The RV value listed here has not yet included the barycentric correction, which is an addition of 13.95 km/s for our observation. Therefore, the true RV measurement for this order is 16.3$\pm$0.4 km/s. Note that the posterior distribution for $R$ is effectively set by our choice of prior (24,800$\pm$1000) on this parameter.}
\end{figure}

We cross-correlate the observed companion spectrum (see Fig.~\ref{fig:tel_cor_spec}) with a model Sonora spectrum that is broadened to the measured instrumental profile ($R$=24,800$\pm$1000). Each spectral order is analyzed separately and the cross-correlation function is defined as:
\begin{equation}
CCF(w) = \frac{\sum_{i=1}^{n} d(i)*m(i-w)}{\sqrt{\sum_{i=1}^{n} d(i)^2}*\sqrt{\sum_{i=1}^{n} m(i-w)^2}},
\label{eq:ccf}
\end{equation}
where $d$ is the observed spectrum, $m$ is the model spectrum broadened by the instrumental resolution $R$, and \textit{w} represents the relative displacement in wavelength space. We sum this quantity over the $n$ points in each spectral order. We calculate the uncertainties $\sigma_i$ on the CCF of the model and data using the jackknife resampling technique:
\begin{equation}
\sigma_{\rm{jackknife}}^2 = \frac{(n-1)}{n} \sum_{i=1}^{n} {(x_i - x)}^2,
\label{eq:jackknife}
\end{equation}
where $x_i$ is the CCF calculated using all but the $i$th AB nod pair, and $x$ is the CCF calculated using all AB nod pairs. The number of samples $n$ is equal to the number of individual AB nod pairs (eight for DH Tau b and two for DH Tau).

We measure the projected rotation rate ($v$sin$i$) and radial velocity (RV) by comparing the measured CCF to a series of model CCFs. Each model CCF is calculated by cross-correlating a model spectrum broadened to the instrumental resolution with that same model additionally broadened by a rotation rate and shifted by an RV offset. The CCF fitting process is illustrated in Fig.~\ref{fig:ccfs}. We fit the CCF using an MCMC framework \texttt{emcee}, implemented by \citet{foreman-mackey_emcee:_2013}. For each MCMC fit, we used 20 walkers and 4000 steps. We removed the first 10\% as burn-in from our resultant chains, which all converged with auto-correlation factors \textgreater 50. In addition to the companion's RV and $v$sin$i$, we include the instrumental resolution $R$ as a third parameter in the MCMC fits. We place a Gaussian prior on $R$ centered at the best fit value of 24,800, with a FWHM equal to the 1$\sigma$ uncertainty of $\pm$1000. Because $R$ is degenerate with $v$sin$i$ in our fits, this prior allows us to account for the effect that uncertainties in $R$ have on our posterior probability distribution for $v$sin$i$. We assume uniform priors for model parameters $v$sin$i$ and RV. Neglecting the constant term, the log likelihood function for our MCMC is then:
\begin{equation}
log{L} = \sum_{i=1}^{n} -0.5(\frac{m_i - d_i}{\sigma_{i}})^2,
\label{eq:loglike}
\end{equation}
where $d$ is the measured CCF, $m$ is the model CCF, and $\sigma_{i}$ is the uncertainty of the CCF at position $i$. For the purpose of these fits we limit the effective range of the CCF to a region centered around the peak with edges at -20 and +25 km/s in units of RV offset (see Fig.~\ref{fig:ccfs}). Given the positive RV shift from the companion, we chose an asymmetric window so that the fitted peak would be symmetric. Our measured $v$sin$i$ is relatively insensitive to the specific window used (within 1$\sigma$). We determined the optimal window region as the one that minimizes the spread in the measured $v$sin$i$ values between different spectral traces.

We measure the $v$sin$i$ and RV of DH Tau b from individual fits to each of the positive and negative traces of orders 1 and 2, resulting in four independent estimates for each parameter. Our measured values are summarized in Table~\ref{tab:vsini_measures}. For our four measurements, we find a reduced $\chi^2$ of 0.9 and 0.3 for $v$sin$i$ and RV, respectively, indicating that the values from these two orders are all consistent within the errors. 

For our final analysis, we carry out a joint fit to the positive and negative traces in orders 1 and 2 using a log likelihood function that is the sum of the individual log likelihood functions from each trace. This gives us a global best-fit RV value of 16.6$\pm$0.3 km/s, and a $v$sin$i$ value of 9.6$\pm$0.7 km/s for the companion.

We repeat this process for the host star spectrum to estimate the instrumental resolution using the known stellar $v$sin$i$ (as described in \S\ref{sec:res}), and measure the RV of the star.

We assess the effect of uncertainties in $T_{\rm{eff}}$ and log($g$) on our measured value for $v$sin$i$ by creating Sonora models with a range of $T_{\rm{eff}}$ and log($g$) values and deriving a new rotation rate for each model. We find that the resulting rotation rates vary by less than 0.5$\sigma$ when we vary $T_{\rm{eff}}$ and log($g$) to values corresponding to the 1$\sigma$ maxima and minima in log($L_{\rm{Bol}}$/$L$\textsubscript{\(\odot\)}) and age ([2400 K, 3.6 m $\rm{s}^{-2}$] and [2200 K, 3.8 m $\rm{s}^{-2}$], respectively), indicating that our rotation measurement is relatively insensitive to our choice of model parameters for the planet. We also test whether the unknown metallicity of DH Tau b influences our spin measurement, we varied the metallicity input to the Sonora models by $\pm$0.5 dex, and repeated our fits with these new models. We found that the resulting spins differed from our solar metallicity spin value by less than 0.9$\sigma$.

\begin{deluxetable}{ccccc}
\tablecolumns{5}
\tabletypesize{\small}
\tablewidth{\textwidth}
\tablecaption{\normalsize{Individual $v$sin$i$ (km/s) and RV (km/s) measurements for DH Tau b} \label{tab:vsini_measures}}
\tablehead{\colhead{Order} & \colhead{$v$sin$i$} & \colhead{$v$sin$i$} & \colhead{RV} & \colhead{RV} \vspace{-8pt} \\
\colhead{number} & \colhead{(Pos)} & \colhead{(Neg)} & \colhead{(Pos)} & \colhead{(Neg)}}
\startdata
\tableline
1 & 10.5$\pm$1.0 & 7.7$\pm$2.5 & 16.5$\pm$0.6 & 16.6$\pm$0.9\vspace{+2pt} \\
2 & 8.6$\pm$1.2 & 9.4$\pm$1.3 & 17.0$\pm$0.6 & 16.3$\pm$0.4\vspace{+2pt} \\
\tableline
1+2 & \multicolumn{2}{c}{9.6$\pm$0.7} & \multicolumn{2}{c}{16.6$\pm$0.3}\\
\enddata
\tablecomments{Measurements of the companion $v$sin$i$ and RV using the positive and negative traces of orders 1 and 2. The error bars represent MCMC fitting uncertainties. The RV values have been corrected for Earth's motion. The last row shows the results of a joint fit to all four traces from orders 1 and 2, which we use as our final measurements.}
\end{deluxetable}

\subsection{Radial Velocities} \label{sec:RV}
For DH Tau, the measured RV is a composite of two effects: the RV of the DH Tau system and the orbital motion of Earth. For DH Tau b, the measured RV also includes the radial component of its orbital velocity. We correct for the Earth's motion (the barycentric correction) with the Python package PyAstronomy\footnote{\href{https://github.com/sczesla/PyAstronomy}{https://github.com/sczesla/PyAstronomy}}, which calculates the relative motion of Earth in the direction of DH Tau at the time of observation. After applying the barycentric correction, which comes out to +13.95 km/s, we find an RV of 16.2$\pm$0.2 km/s for DH Tau, which is consistent with the literature value of 16.52$\pm$0.04 km/s \citep{nguyen_close_2012} at the 1.5$\sigma$ level.

We next calculate the magnitude of the expected orbital motion for DH Tau b. Assuming an edge-on circular orbit with a radius of 320 AU, we find that the RV shift caused by the companion's orbital motion could be as large as 1 km/s. This value corresponds to times when the companion is moving directly towards or away from the observer (i.e., at the time of maximum projected separation between the companion and the star). Given our measured stellar RV of 16.2 km/s, this means that the RV of DH Tau b should be between 15.2-17.2 km/s. We obtain a RV value of 16.6$\pm$0.3 km/s for DH Tau b, which is consistent with the star's radial velocity at the $1.5\sigma$ level. This suggests that our RV measurement is not precise enough to detect the companion's orbital motion. For perspective, a 1 km/s RV signal corresponds to an accuracy of 0.2 pixels for NIRSPEC in $K$ band.

\section{Discussion} \label{sec:discuss}

\subsection{True Rotation Rate and Break-up Velocity}\label{sec:v_b}

\begin{deluxetable*}{lccccccccc}
\tablecolumns{10}
\renewcommand{\arraystretch}{1.4}
\tabletypesize{\footnotesize}
\tablewidth{\textwidth}
\tablecaption{\normalsize{Measured Rotation Rates for Planetary-Mass Companions} \label{tab:object_info}}
\tablehead{\colhead{Name} & \colhead{$v$sin$i$ (km/s)} & \colhead{$P_{\rm{rot}}$ (hr)} & \colhead{$v$ (km/s)} & \colhead{$v$/$v_{\rm{breakup}}$} & \colhead{Spin source} & \colhead{Mass ($M_{\rm{Jup}}$)} & \colhead{Radius ($R_{\rm{Jup}}$}) & \colhead{Age (Myr)} & \colhead{References}}
\startdata
\tableline
DH Tau b & 9.6$\pm$0.7 & n/a & 11.6$\pm$2.7\tablenotemark{c} & $0.12^{+0.03}_{-0.03}$ & This paper & 11-17 & 2.68$\pm$0.22 & 2$\pm$1 & 1,7,8 \\
ROXs 42B b & 9.5$\pm$2.0 & n/a & 11.5$\pm$3.7\tablenotemark{c} & $0.13^{+0.04}_{-0.05}$ & Bryan+2018 & 6-14 & 2.11$\pm$0.11 & 3$\pm$2 & 2,9,10,11 \\
VHS 1256-1257 b & 13.5$\pm$4.0 & n/a & 16.6$\pm$6.4\tablenotemark{c} & $0.12^{+0.07}_{-0.06}$ & Bryan+2018 & 10-21 & 1.11$\pm$0.03 & 225$\pm$75 & 2,12,13 \\
GSC 6214-210 b & 6.1$\pm$4.0 & n/a & 7.7$\pm$5.5\tablenotemark{c} & $0.06^{+0.04}_{-0.05}$ & Bryan+2018 & 15$\pm$2 & 1.91$\pm$0.07 & 11$\pm$2 & 2,14,15 \\
$\beta$ Pic b & 25.0$\pm$3.0 & n/a & 29.6$\pm$7.3\tablenotemark{c} & $0.24^{+0.05}_{-0.07}$ & Snellen+2014 & 13$\pm$3 & 1.47$\pm$0.02 & 23$\pm$3 & 3,16,17,18 \\
2M1207-3932 b & n/a\tablenotemark{b} & 10.7 & 17.7$\pm$1.5 & $0.22^{+0.05}_{-0.05}$ & Zhou+2016 & 5$\pm$2 & 1.38$\pm$0.02 & 10$\pm$3 & 5,19,20 \\
2M0122-2439 b\tablenotemark{a} & n/a\tablenotemark{b} & 6.0 & 20.8$\pm$4.0 & $0.12^{+0.03}_{-0.03}$ & Zhou+2019 & 12-27 & 1.17$\pm$0.02 & 120$\pm$10 & 7,23 \\
AB Pic b\tablenotemark{a} & n/a\tablenotemark{b} & 2.1 & 89.1$\pm$6.0 & $0.70^{+0.05}_{-0.07}$ & Zhou+2019 & 11-14 & 1.40$\pm$0.05 & 15-40 & 7,24 \\
Ross 458 c & n/a\tablenotemark{b} & 6.8 & 19.9$\pm$5.0 & $0.16^{+0.05}_{-0.05}$ & Manjavacas+2019 & 9$\pm$3 & 1.07$\pm$0.02 & 150-800 & 6,21,22 \\
HD 203030 B & n/a\tablenotemark{b} & 7.5 & 21.0$\pm$1.7 & $0.16^{+0.03}_{-0.03}$ & Miles-P\'{a}ez+2019 & 8-15 & 1.19$\pm$0.03 & 30-150 & 25,26 \\
\enddata
\tablenotetext{a}{Marginal detections (2.2-3.1$\sigma$) included in our plots but excluded from the analysis.}
\tablenotetext{b}{Rotation rate calculated using photometric monitoring data.}
\tablenotetext{c}{True rotation rates calculated by dividing out a distribution in sin$i$ from the $v$sin$i$ measurements. For simplicity, the $v$ uncertainties listed are averages of upper and lower errors.}
\tablecomments{A summary of spin measurements for planetary-mass companions from detections of rotational line broadening (top five) and photometric periodicity (bottom five). For the rotational broadening measurements, we list the directly measured quantity $v$sin$i$ as well as the derived $v$ estimate, calculated by dividing out a distribution of sin$i$ from $v$sin$i$. For the photometric periods, we list the directly measured $P_{\rm{rot}}$, as well as the calculated $v$. For each object, $v_{\rm{breakup}}$ is computed from its mass and radius estimates, and used to calculate the ratio $v$/$v_{\rm{breakup}}$.}
\tablerefs{(1) \citet{itoh_young_2005}, (2) \citet{bryan_constraints_2018}, (3) \citet{snellen_fast_2014}, (4) \citet{zhou_discovery_2016}, (5) \citet{manjavacas_ross_2019}, (6) \citet{zhou_cloud_2019}, (7) \citet{luhman_discovery_2006}, (8) \citet{bertout_evolution_2007}, (9) \citet{kraus_three_2013}, (10) \citet{currie_direct_2013}, (11) \citet{bowler_spectroscopic_2014}, (12) \citet{gauza_discovery_2015}, (13) \citet{stone_adaptive_2016}, (14) \citet{ireland_two_2010}, (15) \citet{lachapelle_characterization_2015}, (16) \citet{lagrange_giant_2010}, (17) \citet{dupuy_model-independent_2019}, (18) \citet{mamajek_age_2014}, (19) \citet{chauvin_giant_2005}, (20) \citet{bell_self-consistent_2015}, (21) \citet{goldman_new_2010}, (22) \citet{scholz_hip_2010}, (23) \citet{bowler_planets_2013}, (24) \citet{chauvin_companion_2005}, (25) \citet{miles-paez_cloud_2019}, (26) \citet{miles-paez_prototypical_2017}}
\end{deluxetable*}

In the absence of any braking mechanism, we would expect DH Tau b to spin up to near the predicted break-up velocity $v_{\rm{breakup}}$ as it accretes gas and acquires angular momentum from the circumplanetary disk. The present-day ratio of $v$/$v_{\rm{breakup}}$ therefore provides a measure of the efficiency of angular momentum loss mechanisms both during and after the end of accretion. We convert our $v$sin$i$ measurement for DH Tau b to a distribution in $v$ by dividing by a probability distribution of the form sin$i$, which we generate assuming a uniform distribution in cos$i$. This yields an estimate of the true rotation rate $v$ = $11.6^{+2.5}_{-2.9}$ km/s. The break-up velocity is calculated by equating the gravitational force and the centripetal force at the surface of the object
\begin{equation}
    v_{\rm{breakup}} = \sqrt{GM/R}
\end{equation}
where $M$ and $R$ are the mass and radius of DH Tau b. We estimate the radius using COND evolutionary models \citep{baraffe_evolutionary_2003}, which require age and luminosity measurements. Assuming an age of 2$\pm$1 Myr, and using a log($L_{\rm{Bol}}$/$L$\textsubscript{\(\odot\)}) of -2.71$\pm$0.12 as measured by \citet{luhman_discovery_2006}, we use a Monte Carlo sampling approach to generate a distribution for radius by drawing random pairs of age and luminosity and interpolating the COND models to infer the corresponding radius. In this manner, we estimate a radius of $2.68^{+0.21}_{-0.22}$ $R_{\rm{Jup}}$ for DH Tau b. We also apply the sampling approach to come up with a new mass estimate of $14.2^{+2.4}_{-3.5}$ $M_{\rm{Jup}}$ for DH Tau b, which is consistent with the previous measurement of $8-22$ $M_{\rm{Jup}}$ from \citet{luhman_discovery_2006}, who used models from \citep{chabrier_deuterium_2000} and \citep{burrows_1997}. We use our own mass estimate for DH Tau b in the analysis for the sake of consistency. Our mass and radius estimates give a break-up velocity of $97^{+9}_{-13}$ km/s for DH Tau b. This suggests that the companion is rotating at a significantly lower rate than its breakup velocity ($v$/$v_{\rm{breakup}}=0.12\pm0.03$). 

\begin{figure}[t!]
    \centering
    \includegraphics[width=\linewidth]{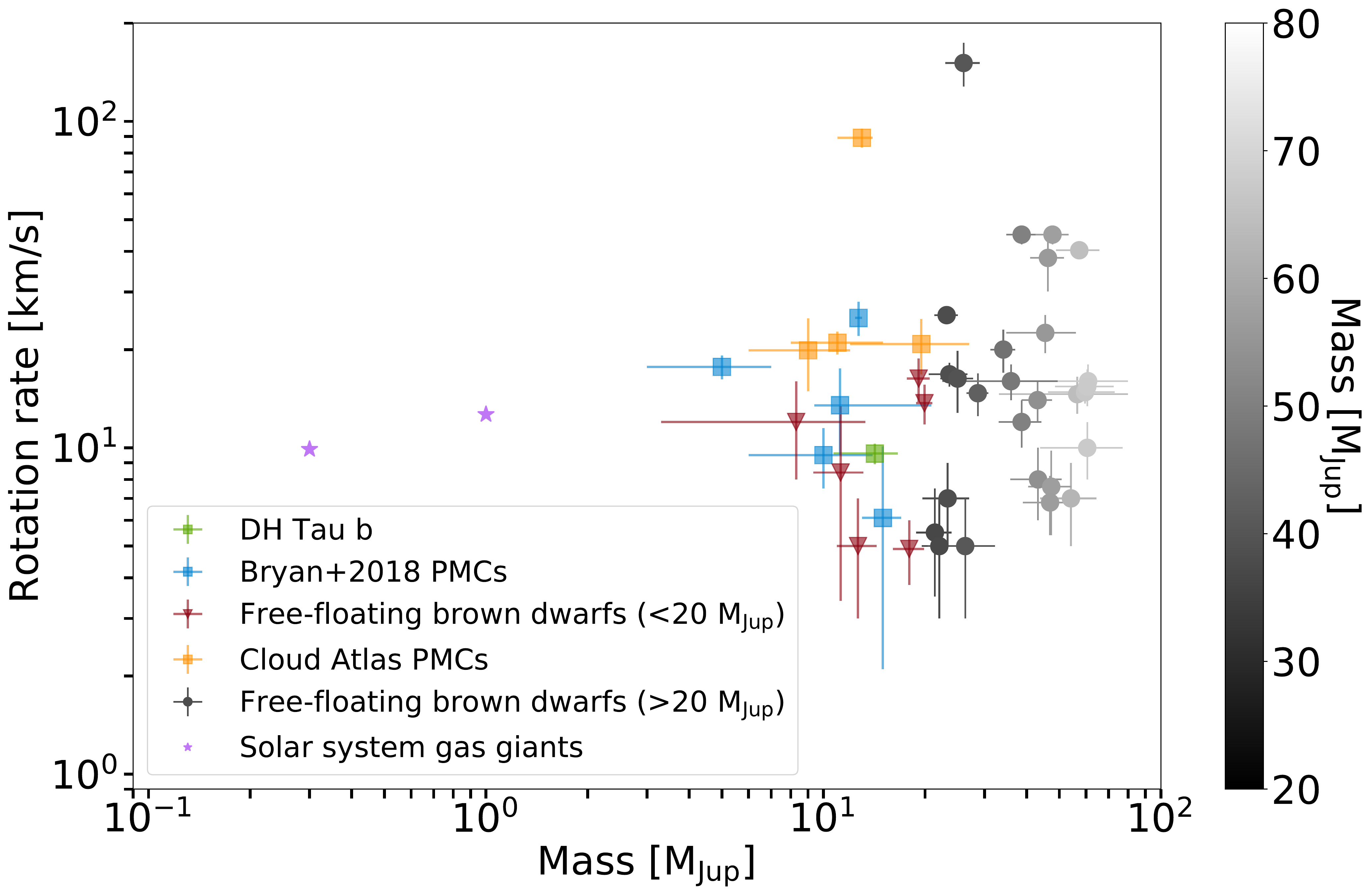}
    \caption{Measured rotational velocity as a function of mass for PMCs, solar system gas giants, and free-floating brown dwarfs with $M$\textless20 $M_{\rm{Jup}}$. We also show more massive brown dwarfs ($20-80$ $M_{\rm{Jup}}$) as filled grey circles, where the shading indicates the mass. The samples includes five bound planetary mass companions (blue squares) and six free-floating brown dwarfs (red triangles) from \citet{bryan_constraints_2018}, four additional bound companions from \citet{zhou_cloud_2019} and \citet{manjavacas_ross_2019} (yellow squares), the solar system gas giants (purple stars), and our new rotation rate measurement for DH Tau b (green square).
    \label{fig:mass_plot}}
\end{figure}

\subsection{Rotation Rate as a Function of Mass}

In Fig. \ref{fig:mass_plot}, we explore the observed trend between mass and rotation described in \citet{bryan_constraints_2018} by adding our new $v$sin$i$ measurement for DH Tau b, as well as newly published rotation rates for four bound companions AB Pic b, 2M0122 B, Ross 458 b, and HD 203030 B from \citet{zhou_cloud_2019}, \citet{manjavacas_ross_2019}, and \citet{miles-paez_cloud_2019}. The latter four data points are equatorial rotation rates derived from photometric monitoring data from an ongoing \emph{Hubble Space Telescope} program \citep[`Cloud Atlas,' PI D. Apai; see summary by][]{manjavacas_cloud_2019}. We convert rotation periods to velocities with radius estimates which we derive from COND models \citep{baraffe_evolutionary_2003}. Although we show equatorial rotation rates for AB Pic b and 2M0122 B in our figures, we do not include these two objects in our subsequent analysis as their photometric variability is detected with marginal significances of 2.2-3.1$\sigma$, calculated with the assumption that the intrinsic shape of the lightcurve is a single component sinusoid \citep{zhou_cloud_2019}. For context, we also plot the rotation rates of higher mass brown dwarfs ($20-80$ $M_{\rm{Jup}}$) with measured rotation rates from the literature \citep{joergens_uves_2001,white_very_2003,zapatero_osorio_photometric_2003,zapatero_osorio_dynamical_2004,mohanty_2005,kurosawa_radial_2006-1,rice_physical_2010-1,cody_precision_2010}.

In the following discussion, we denote `planetary-mass objects' as the combined sample of bound PMCs, free-floating planetary-mass brown dwarfs ($M$\textless20 $M_{\rm{Jup}}$), and solar system gas giants. We exclude the solar system ice giants and terrestrial planets from this sample because unlike gas giants, which are characterized by an extensive gas accretion phase, these planets have distinct spin evolutions that are dominated by the accretion of solids and further altered by tides and collisions \citep{Correia_2001, Morbidelli_2012}. \cite{bryan_constraints_2018} concluded that there is no evidence for a correlation between mass and rotation rate in the planetary-mass regime. We quantify the effect of the new measurements of DH Tau b and Cloud Atlas PMCs by calculating an updated Pearson's coefficient of 0.03 between mass and rotation rate for our sample of planetary-mass objects. This finding suggests that the efficiency of the spin regulation mechanism is not sensitive to object mass for companions and free-floating brown dwarfs with masses between $0.3-20$ $M_{\rm{Jup}}$.

\subsection{Angular Momentum Evolution}
In Fig.~\ref{fig:age_plot}, we plot the rotation rates normalized by break-up velocity as a function of age for the same sample in order to search for evidence of angular momentum evolution. To calculate $v_{\rm{breakup}}$ for the other objects, we estimate their radii using the same Monte Carlo sampling approach described in \S\ref{sec:v_b}, which takes the luminosity and age as inputs. We use luminosity and age values from literature. For the masses, we use literature values for all objects except DH Tau b (see \S\ref{sec:v_b}). The mass and radius values adopted for the PMCs among the sample are listed in Table~\ref{tab:object_info}.

We find that the measured $v$/$v_{\rm{breakup}}$ for DH Tau b is consistent with the average value for the five bound companions in \citet{bryan_constraints_2018}. We compare the ratio $v$/$v_{\rm{breakup}}$ of our updated sample of eight PMCs (DH Tau b, ROXs 42B b, VHS 1256-1257 b, GSC 6214-210 b, $\beta$ Pic b, 2M1207-3932 b, Ross 458 c, and HD 203030 B) with the $v$/$v_{\rm{breakup}}$ of the sample of six free-floating brown dwarfs with $M$\textless 20 $M_{\rm{Jup}}$ (OPH 90, USco J1608-2315, PSO J318.5-22, 2M0355+1133, and KPNO Tau 4) from \citet{bryan_constraints_2018} using a two-sample Anderson-Darling test, which tests the null hypothesis that two samples are drawn from the same population. We use the Anderson-Darling test for our comparison instead of an error-weighted average since it is not skewed by small error sizes, and additionally takes into account the intrinsic scatter in the measurements and the small sample size. We find a p-value of 0.47 (0.6$\sigma$), indicating that our data are consistent with the null hypothesis that the $v$/$v_{\rm{breakup}}$ values for bound and free-floating objects come from the same distribution.

\begin{figure}[t!]
    \centering
    \includegraphics[width=\linewidth]{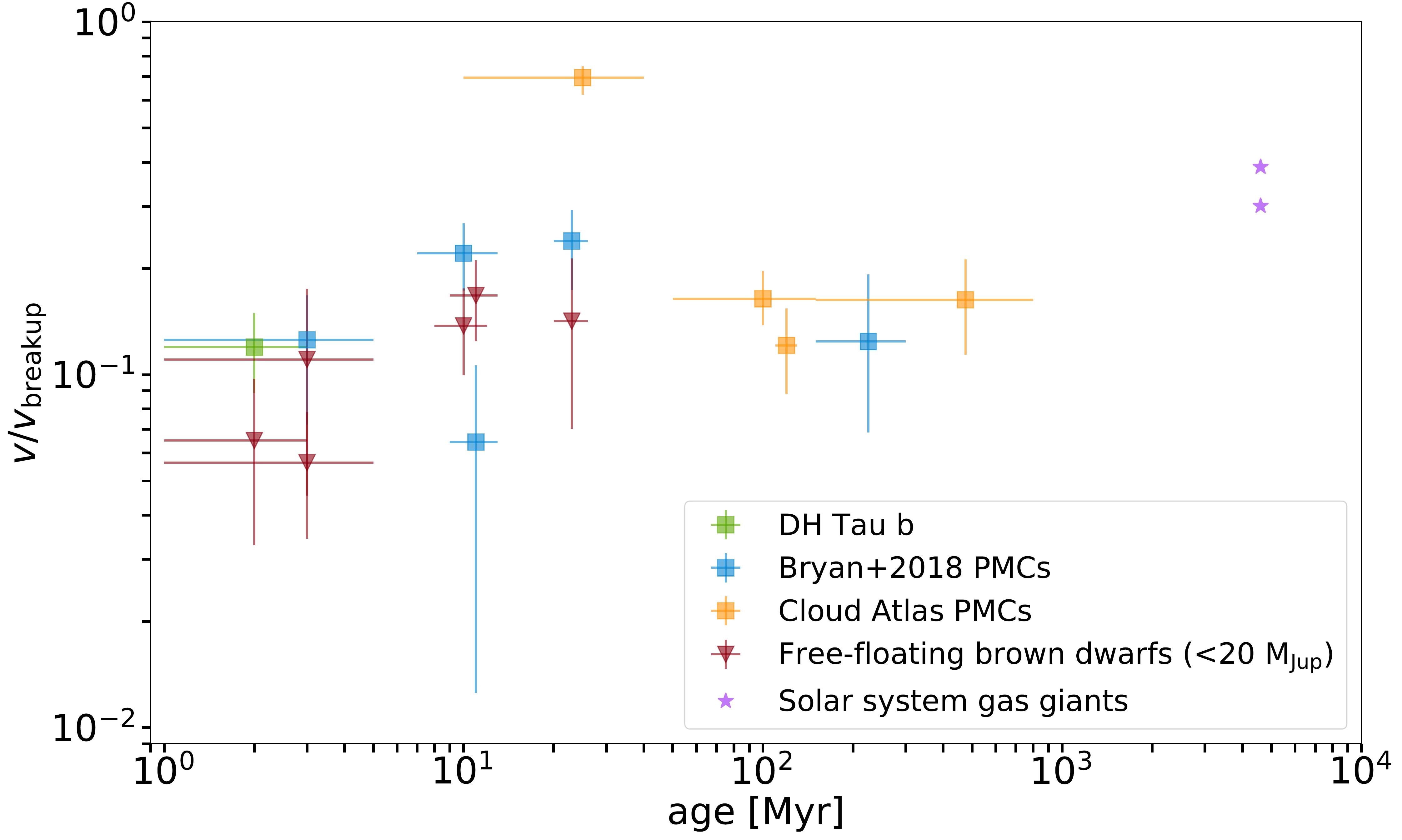}
    \caption{Evolution of rotation rate as a fraction of breakup velocity. We show the five PMCs (blue squares) and six free-floating planetary-mass brown dwarfs (red triangles) from \citet{bryan_constraints_2018}, DH Tau b (green square), as well as Jupiter and Saturn (purple squares). We also include four \textit{Cloud Atlas} PMCs (Ross 458 b, AB Pic b, 2M0122 b, and HD 203030 B) with photometric rotation rotation periods. The uncertainties in $v$/$v_{\rm{breakup}}$ include uncertainties in the measured value for $v$sin$i$, the estimated break-up velocity $v_{\rm{breakup}}$, and assume a random orientation for the spin axis $i$ (for the objects with $v$sin$i$).
    \label{fig:age_plot}}
\end{figure}

We quantify the angular momentum evolution of the bound PMCs and free-floating brown dwarfs with masses less than 20 $M_{\rm{Jup}}$ by calculating a Pearson's correlation coefficient between their age and $v$/$v_{\rm{breakup}}$ values. We find a coefficient of 0.16, indicating that the observations do not reveal any significant angular momentum evolution. We exclude Jupiter and Saturn in this calculation because they both have significantly lower masses and smaller planet-star separations than the other objects in our sample, and it is therefore unclear whether or not they formed via the same mechanism. The fact that the PMCs and brown dwarfs have similarly low values for $v$/$v_{\rm{breakup}}$ (between $6-24\%$), and that neither population appreciably changes its angular momentum over the several hundred Myrs after the end of accretion, is in good agreement with a scenario in which these objects shed most of their primordial angular momentum by magnetic coupling to a circumplanetary gas disk. As discussed in \S\ref{sec:intro}, \citet{batygin_terminal_2018} suggests that magnetic coupling between a faster-rotating planet and a slower-rotating circumplanetary disk might provide an efficient braking mechanism for this $0.3-20$ $M_{\rm{Jup}}$ population. In this model, the disk extracts angular momentum from the planet, while meridional circulation of gas within the Hill sphere recycles this angular-momentum rich gas back into the circumstellar nebula, thereby decreasing the planet's rotation rate. In fact, there is indirect evidence that DH Tau b is actively accreting, as this object has both H$\alpha$ and Pa$\beta$ emission lines in its spectrum \citep{zhou_accretion_2014, bonnefoy_library_2014, wolff_upper_2017}. This accretion would presumably be mediated via a circumplanetary gas disk, but \citet{wu_explanation_2017} did not find any evidence for such a disk when they observed this object with ALMA. However, they note that a compact and optically thick disk could lie below the detection threshold of their ALMA observations. 

In this scenario, the similar rotation rates of bound and isolated planetary-mass objects would suggest that both populations have broadly similar circumplanetary disk properties, or that the magnetic coupling mechanism is relatively insensitive to specific disk properties, since only the inner edge of the disk matters for magnetic coupling \citep{batygin_terminal_2018}.

The lack of angular momentum evolution for our sample of planetary-mass objects ($M$\textless 20$M_{\rm{Jup}}$) is in contrast with observed trends for more massive brown dwarfs ($M$\textgreater 20$M_{\rm{Jup}}$) and stars \citep[e.g.,][]{bouvier_angular_2014}. While stars shed substantial amounts of angular momentum later in their lifetimes via magnetized stellar winds, the handful of studies that have explored angular momentum evolution in the substellar ($20-80$ $M_{\rm{Jup}}$) regime have found that brown dwarfs spin down more slowly than stars. If similar spin regulating mechanisms operate in the substellar mass regime, they operate with less efficiency \citep[e.g.,][]{zapatero_osorio_spectroscopic_2006, scholz_rotation_2015}. Our conclusion that no significant angular momentum evolution occurs in the planetary-mass regime might therefore be a reasonable extension of this trend.

While we do not include AB Pic b in our analysis due to the marginal significance of its photometric variability detection, we note that the estimated rotation rate for this object would make it the fastest spinning young planetary-mass object currently known ($v$/$v_{\rm{breakup}}$ = 0.70, assuming a radius of 1.4 $R_{\rm{Jup}}$). As noted by \citet{zhou_cloud_2019}, if AB Pic b shrinks to R = 1 $R_{\rm{Jup}}$ while conserving angular momentum, it would attain a velocity closely approaching breakup. Based on its estimated age of $10-40$ Myr we would expect AB Pic b's circumplanetary disk to have dispersed, preventing it from shedding this excess angular momentum via disk coupling. This suggests either that AB Pic b had a circumplanetary disk with properties that differed appreciably from those of the other objects in our sample, therefore preventing it from effectively shedding angular momentum during the final stages of accretion, or its rotation period is underestimated. As discussed in \citet{zhou_cloud_2019}, the rotation period might be underestimated if the measured photometric modulations are dominated by higher order planetary-scale waves, which would cause the observed period to be a higher harmonic of the full rotation period.

\begin{figure}[t!]
    \centering
    \includegraphics[width=\linewidth]{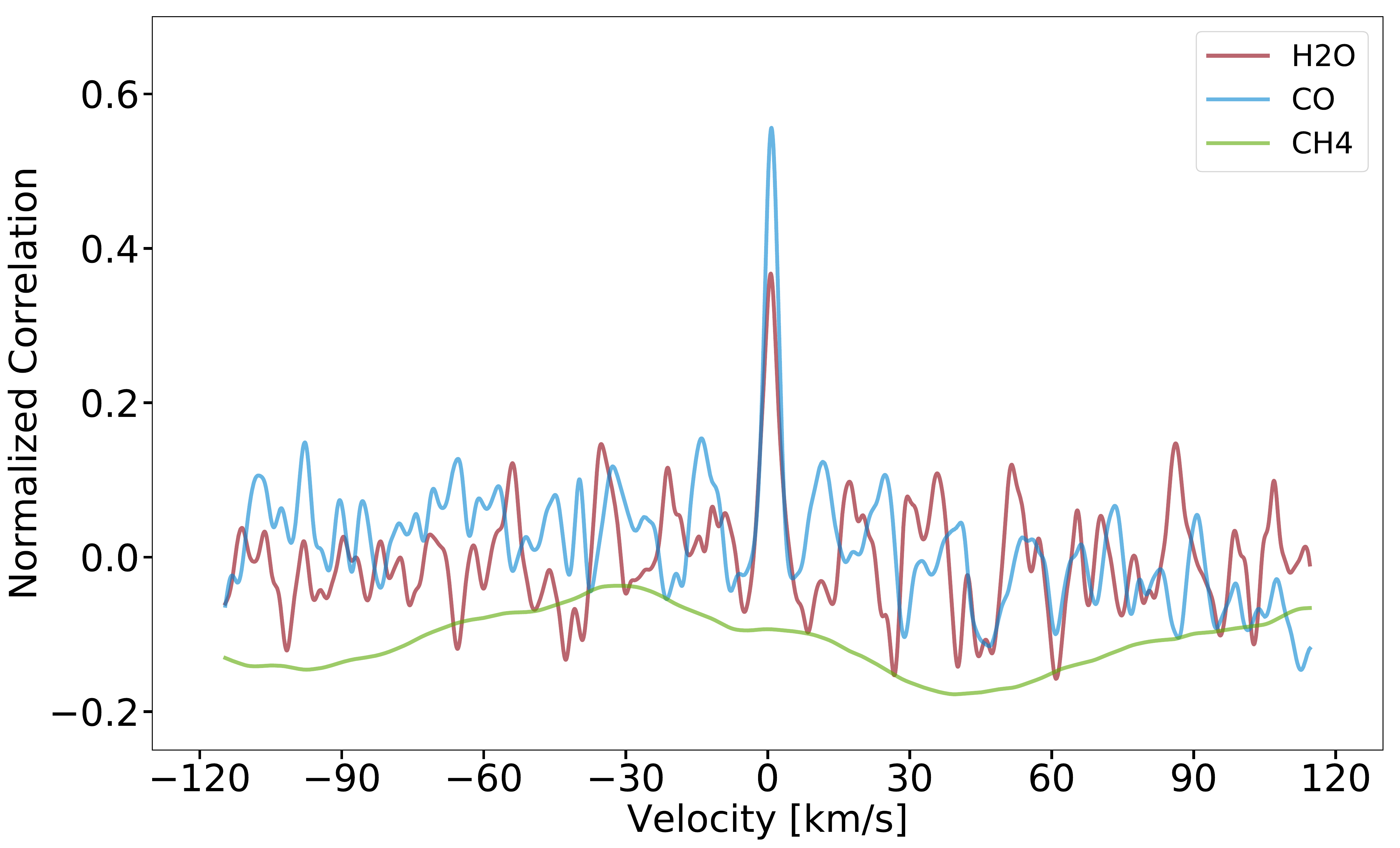}
    \caption{Cross-correlation of DH Tau b's spectrum with individual molecular species including water, carbon monoxide, and methane. The CCF is calculated from the combined spectrum of orders 1 and 2 using both the positive and negative traces. Water and carbon monoxide are clearly detected in the spectrum, but we find no evidence for the presence of methane.
    \label{fig:atm_comp}}
\end{figure}

\subsection{Atmospheric Composition}
We cross-correlate our spectrum for DH Tau b with models that include collision-induced opacity of hydrogen and helium plus one additional molecule (H$_2$O, CO, or CH$_4$) in order to determine the main molecular species in the atmosphere of DH Tau b. These models use the same pressure--temperature profiles and abundances of the included species as standard Sonora models with $T_{\rm eff}$=2300 K, log($g$)=3.7, and [M/H]=0.0. The opacities from all other species are removed to recalculate the single-species spectrum.

As shown in Fig.~\ref{fig:atm_comp}, we detect water and carbon monoxide with signal-to-noise ratios (SNR) of 5.9 and 8.7, respectively. We define CCF SNR here as the ratio between the CCF peak height and RMS of the CCF wings. We do not detect methane, but this is expected given the relatively high effective temperature of DH Tau b ($\sim$2300 K).  At these temperatures, methane reacts with water to produce carbon monoxide and molecular hydrogen, resulting in a CO-dominated carbon chemistry. Objects with effective temperatures above a threshold temperature of $\sim$1200-1300 K are expected to have atmospheres abundant in CO with little CH$_4$ \citep{Kirkpatrick_new_2005,Lodders_chemistry_2006}.

\subsection{Additional Probes of DH Tau b's Formation History}\label{sec:future_work}
In principle, it should be possible to directly fit DH Tau b's spectrum in order to determine the relative abundances of key molecular species including water, carbon monoxide, carbon dioxide, and methane. These molecular abundances could then be converted into an elemental C/O ratio using chemical models. By comparing the C/O ratio of DH Tau b to that of its host star, we could place constraints on potential formation scenarios  \citep[see e.g.,][]{konopacky_detection_2013}. However, in order to reliably extract molecular abundances we must first improve our understanding of the nature of noise in our spectra, which includes wavelength-correlated structures from imperfect removal of telluric lines that might bias retrieval results. 

Alternatively, we might instead seek to constrain the orbital and spin geometry of the DH Tau system by combining relative astrometry and rotation measurements, in a similar manner as \citet{bowler_young_2017} did for the planetary-mass companion ROXs 12 B. Several studies have been able to detect orbital motion for planetary-mass companions at wide separations and constrain orbital parameters \citep{bryan_searching_2016, bowler_young_2017, pearce_orbital_2018}. If a photometric rotation rate and rotational line broadening measurement are additionally available for the host star, one can then constrain the orbital inclination of the companion with respect to the spin axis of the star. There are published measurements for both quantities in the literature for DH Tau \citep{bouvier_coyotes_1995,nguyen_close_2012}. While 9 epochs of astrometry spanning 19 years exist for DH Tau b, \citet{bowler_population_2019} found large discrepancies in the relative astrometry of DH Tau b between different instruments, precluding a constraint on the companion's orbit at this time. Nonetheless, if we find the planet's orbit is misaligned with respect to the star's spin axis, it would suggest that DH Tau b might have formed via turbulent fragmentation, or in a protoplanetary disk that was torqued by DI Tau, a stellar companion 2210 AU away from the DH Tau system. It is also possible that DH Tau b formed in an aligned disk but was later dynamically excited by interactions with another close-in companion, although \citet{bryan_searching_2016} find that this scenario is unlikely.

\section{Conclusions} \label{sec:conclude}
We obtain a high-resolution $K$ band spectrum for the planetary-mass companion DH Tau b and its host star using the near-infrared spectrograph NIRSPEC at the Keck II telescope. We measure a projected rotation rate $v$sin$i$ of 9.6$\pm$0.7 km/s for DH Tau b, which converts to a rotational velocity $v$ of $11.6^{+2.5}_{-2.9}$ km/s after taking into account the distribution of spin inclinations $i$.  We conclude that this relatively young ($\sim$2 Myr) object is most likely rotating at just 12$\pm$3\% of its break-up velocity, indicating that it was able to effectively shed most of its primordial angular momentum prior to the end of accretion.  This observation is in good agreement with models of magnetic coupling between the planet and its circumplanetary disk \citep{takata_despin_1996, batygin_terminal_2018}, which predict that this mechanism should provide an efficient means of angular momentum dissipation. Interestingly, previous studies have found active accretion signatures for DH Tau b, suggesting that it may still possess a cicumplanetary gas disk \citep{zhou_accretion_2014, bonnefoy_library_2014, wolff_upper_2017}. 

We compare our spin measurement for DH Tau b to those of nine other planetary-mass companions from the literature \citep{bryan_constraints_2018, manjavacas_ross_2019, zhou_cloud_2019, miles-paez_cloud_2019}, as well as a sample of six free-floating brown dwarfs with similar masses from \citet{bryan_constraints_2018}. We find that the rotation rate distribution for both samples are consistent with being drawn from the same underlying population.  We plot the measured rotation rates for these objects as a function of companion mass and system age and find no evidence for a correlation with either parameter. Our findings suggest that either both populations of objects formed via the same mechanism, or that both had broadly similar disk properties despite differing formation mechanisms.

\acknowledgments
The data presented herein were obtained at the W. M. Keck Observatory, which is operated as a scientific partnership among the California Institute of Technology, the University of California, and the National Aeronautics and Space Administration. The Observatory was made possible by the generous financial support of the W. M. Keck Foundation. The authors wish to recognize and acknowledge the very significant cultural role and reverence that the summit of Maunakea has always had within the indigenous Hawaiian community. We are most fortunate to have the opportunity to conduct observations from this mountain. This work was funded, in part, by a Summer Undergraduate Research Fellowship (SURF) from California Institute of Technology. MLB is supported by the Heising-Simons Foundation 51 Pegasi b Fellowship. BPB acknowledges support from the National Science Foundation grant AST-1909209.

\facilities{KeckII/NIRSPEC}
\software{\texttt{emcee}~\citep{foreman-mackey_emcee:_2013}, \texttt{PyAstronomy}~(\url{https://github.com/sczesla/PyAstronomy})}

\bibliography{DHTaub}
\bibliographystyle{aasjournal}

\end{document}